\documentclass[onecolumn]{aa}
\usepackage{graphicx}
\usepackage{txfonts}

\usepackage{epsfig}

\usepackage{psfrag}

\newcommand{\moa}{MOA~2002-BLG-33}
\newcommand{\etal}{et al.}
\newcommand{\chisq}{$\chi^{2}$}

\begin{document}

\title{Determination of stellar shape in microlensing event \moa}
 
\author{N.J. Rattenbury\inst{1} \and F. Abe\inst{2} \and D.P. Bennett\inst{3} \and I.A. Bond\inst{4} \and J.J. Calitz\inst{5} \and A. Claret\inst{6} \and K.H. Cook\inst{7} \and 
   Y. Furuta\inst{2} \and A. Gal-Yam\inst{8} \and J-F. Glicenstein\inst{10} \and J.B. Hearnshaw\inst{11} \and P.H. Hauschildt\inst{12} \and  
   P.M. Kilmartin\inst{11} \and Y. Kurata\inst{2} \and K. Masuda\inst{2} \and D. Maoz\inst{9} \and Y. Matsubara\inst{2} \and P.J. Meintjes\inst{5} \and M. Moniez\inst{13} \and
   Y. Muraki\inst{2} \and S. Noda\inst{2} \and E.O. Ofek\inst{9} \and K. Okajima\inst{2} \and L. Philpott\inst{14} \and S.H. Rhie\inst{3} \and
   T. Sako\inst{2} \and D.J. Sullivan\inst{15} \and T. Sumi\inst{16} \and D.M. Terndrup\inst{17} \and P.J. Tristram\inst{11} \and J. Wood\inst{14} \and T. Yanagisawa\inst{2} \and P.C.M. Yock\inst{14}}

   \institute{
Jodrell Bank Observatory, Macclesfield, Cheshire SK11 9DL, UK
\and
   Solar Terrestrial Environment Laboratory, Nagoya University, Nagoya 464-8601, Japan
   \and
   Department of Physics, Notre Dame University, Notre Dame, IN 46556, USA
   \and
    Institute of Information and Mathematical Sciences, Massey University, Auckland, New Zealand
   \and
   Department of Physics, University of the Free State, Bloemfontein, South Africa
   \and
   Instituto de Astrof\'{i}sica de Andaluc\'{i}a, CSIC, Apartado 3004, 18080 Granada, Spain 
   \and
   Institute for Geophysics and Planetary Physics, Lawrence Livermore National Laboratory, CA 94563, USA
   \and
   Hubble Fellow, Caltech Astronomy, MS 105-24, California Institute of Technology, 1200 E. California Bl., Pasadena, California
   \and
School of Physics and Astronomy, Tel-Aviv University, Tel-Aviv 69978, Israel 
\and
   CEA, DSM, DAPNIA, Centre d'\'{E}tudes de Saclay, F-91191 Gif-sur-Yvette Cedex, France
   \and
   Department of Physics and Astronomy, University of Canterbury, Private Bag 4800, Christchurch, New Zealand
   \and
   Universit\"{a}t Hamburg, Sternwarte, Gojenbergsweg 112, D-21029 Hamburg, Germany
   \and
   Laboratoire de l'Acc\'{e}l\'{e}rateur Lineaire, IN2P3 CNRS, Universit\'{e} de Paris-Sud, F-91405 Orsay Cedex, France
   \and
   Faculty of Science, University of Auckland, Private Bag 92019, Auckland, New Zealand
   \and
   School of Chemical and Physical Sciences, Victoria University, PO Box 600, Wellington, New Zealand
   \and
   Department of Astrophysical Sciences, Princeton University, Princeton NJ 08544, USA
   \and
   Department of Astronomy, Ohio State University, Columbus OH 43210, USA
   }

\abstract{

We report a measurement of the shape of the source star in microlensing event \moa. The lens for this event was a close binary whose centre-of-mass passed almost directly in front of the source star. At
this time, the source star was closely bounded on all sides by a caustic of the lens. This allowed the oblateness of the source star to be constrained. We found that $a/b = 1.02^{+0.04}_{-0.02}$ where $a$ and $b$ are its semi-major and semi-minor axes respectively. The angular resolution of this measurement is approximately 0.04 $\mu\rm{arcsec}$. We also report HST images of the event that confirm a previous identification of the source star as an F8-G2 turn-off  main-sequence star.}

\maketitle

\section{Introduction} 
The oblateness of rotating stars has recently been measured using optical interferometry. The apparent shapes of Altair and Vega were reported by van Belle \etal\ (2001) using the Palomar Testbed Interferometer; and that of Achernar by Domiciano de Souza \etal\ (2003) using the VLTI. 

Further high accuracy observations of the shapes of rotating stars are required in order to test aspects of stellar models, such as gravity darkening (von Zeipel, 1924). 

Oblateness measurements have not been reported for other stars, partly because of the difficulty of carrying out the measurements, and partly because most stars are assumed to be spherical. Here we present a direct measurement of the apparent oblateness of a distant, dwarf star obtained by gravitational microlensing.

Gravitational microlensing occurs when light from a distant ``source'' star is deflected by the gravitational field of an intervening ``lens'' star, as predicted by Einstein. The lens star produces a magnified and distorted image of the source star. Because all stars are in motion, any significant magnification is transitory, occurring on time-scales of weeks to months. The magnification can be measured from the observed luminosity of the source star as the lens-source alignment changes. This yields a ``light curve'' from which information on both the lens and the source stars can be obtained. 

Previously we reported a microlensing event, \moa, where a distant solar-like star was lensed by a binary star (Abe \etal, 2003). The light curve was measured in two passbands (infrared and visual) and used to determine the characteristics of the binary lens system, and also the limb-darkening on the source star. In that analysis it was assumed that the source star was spherical. However, if a star is rotating sufficiently rapidly, it may present an elliptical aspect. Here we present a more general analysis in which possible ellipticity of the source star profile is allowed for.

\section{\moa}

The microlensing event \moa\ (hereafter MOA-33) was remarkable for several reasons. The lens was a  close binary whose centre-of-mass was almost co-linear with the source star, as seen from Earth,  at one stage during the event, resulting in a very high magnification of the source star. During the time of this high intrinsic magnification, the source star passed through the small central ``caustic'' of the binary lens\footnote{In gravitational lensing with binary or more complex lenses, ``caustic'' curves (closed curves) form on which the magnification is extremely high. As the source passes through caustic lines, abrupt but continuous changes in source flux are observed.}.

\begin{figure}
\begin{center}
\psfrag{xlabel}{\LARGE{$\!R'_{\rm{E}}$}}
\psfrag{ylabel}{\LARGE{$R'_{\rm{E}}$}}
\psfrag{times}{\large{$\times 10^{-3}$}}
\psfrag{First contact}{}
\psfrag{Last contact}{}
\psfrag{2460.15}{}
\psfrag{2460.80}{}
\psfrag{beta}{\LARGE{$\beta$}}
\psfrag{a}{\LARGE{$a$}}
\psfrag{b}{\LARGE{$b$}}
\psfrag{T1}{}
\psfrag{T2}{}
\psfrag{T3}{}
\psfrag{eq}{\LARGE{$e = \sqrt{1 - \frac{b^{2}}{a^{2}}}$}}

\scalebox{0.70}{\includegraphics{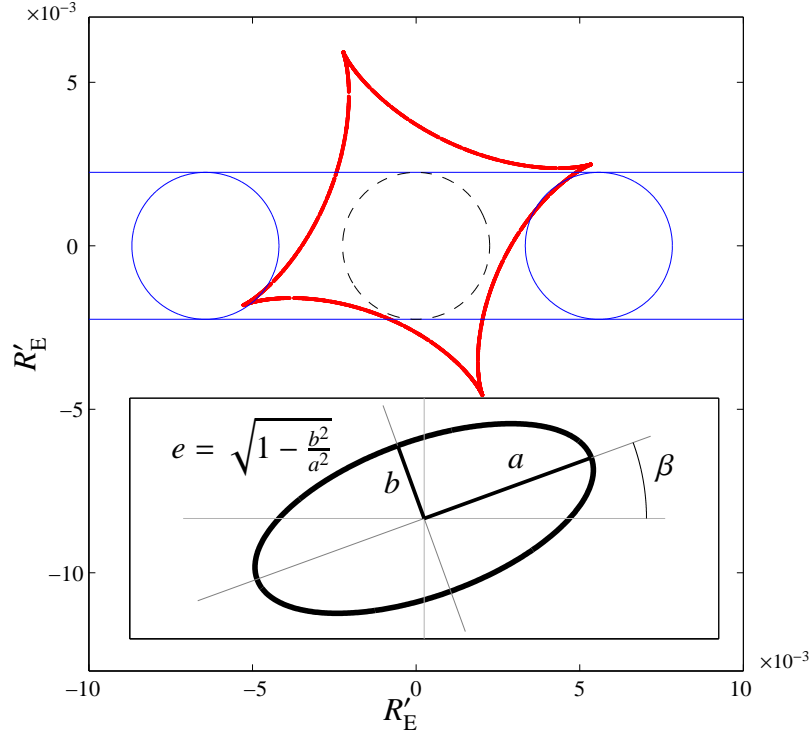}}

\caption[]{The caustic curves (red) of the binary lens of event MOA-33. The source star (blue), shown to scale, has a relative motion horizontally from left to right. The axes are in units of the Einstein ring projected to the source plane, i.e. $R'_{\rm{E}} = R_{\rm{E}} {D_{\rm{OS}}}/{D_{\rm{OL}}}$ where $R_{\rm{E}}$ is the Einstein ring radius and $D_{\rm{OS}}$ and $D_{\rm{OL}}$ are the distances to the source and lens stars respectively. The Einstein ring radius is $R_{E} = \sqrt{\frac{4GM}{c^{2}} D_{\rm{OS}}(1-f)f}$, where $f = D_{\rm{OL}} /  D_{\rm{OS}}$ and $M$ is the total mass of the lens. The circular source star profile is shown at the moments of the caustic entry and exit, corresponding to rapid changes in magnification.  During a finite time interval the caustic curve completely enclosed the source star (dashed line). Any significant deviation from a circular source profile will produce significant features in the light curve. Inset: the current work uses an elliptical source profile, with apparent eccentricity, $e$, and position angle $\beta$ with respect to the source star track.}

\label{fig:caustic}
\end{center}
\end{figure}

Event MOA-33 was monitored by several collaborating groups, and the light curve was well sampled with good accuracy through most of the event. Most importantly for the current work, the source star was at one stage during the event neatly bounded by the central caustic. Figure~\ref{fig:caustic} shows the central caustic due to the binary lens and the circular source star used to model event MOA-33 by Abe~\etal\ (2003). At event time $HJD-2450000 \simeq 2460.5$, the source star was closely bounded by the caustic curve. Any significant excursion from a circular source star profile will be apparent due to this unique situation. 

The photometry of the source star reported in Abe~\etal\ (2003) yielded a dereddened colour $V-I = 0.74 \pm 0.10$ and a dereddened magnitude $m_{\rm{I}} = 16.78\pm 0.10$. These values correspond to an F8 - G2 main-sequence dwarf star with $B-V=0.57\pm0.04$ at a distance of 4-6 kpc (Girardi~\etal, 2002). They are also consistent with a later G-type main-sequence star as close as 3 kpc, but such a star is less likely to be gravitationally lensed.

\section{Source star identification by the HST}
\begin{figure}
\begin{center}
\begin{tabular}{cc}
\scalebox{0.4275}{\includegraphics{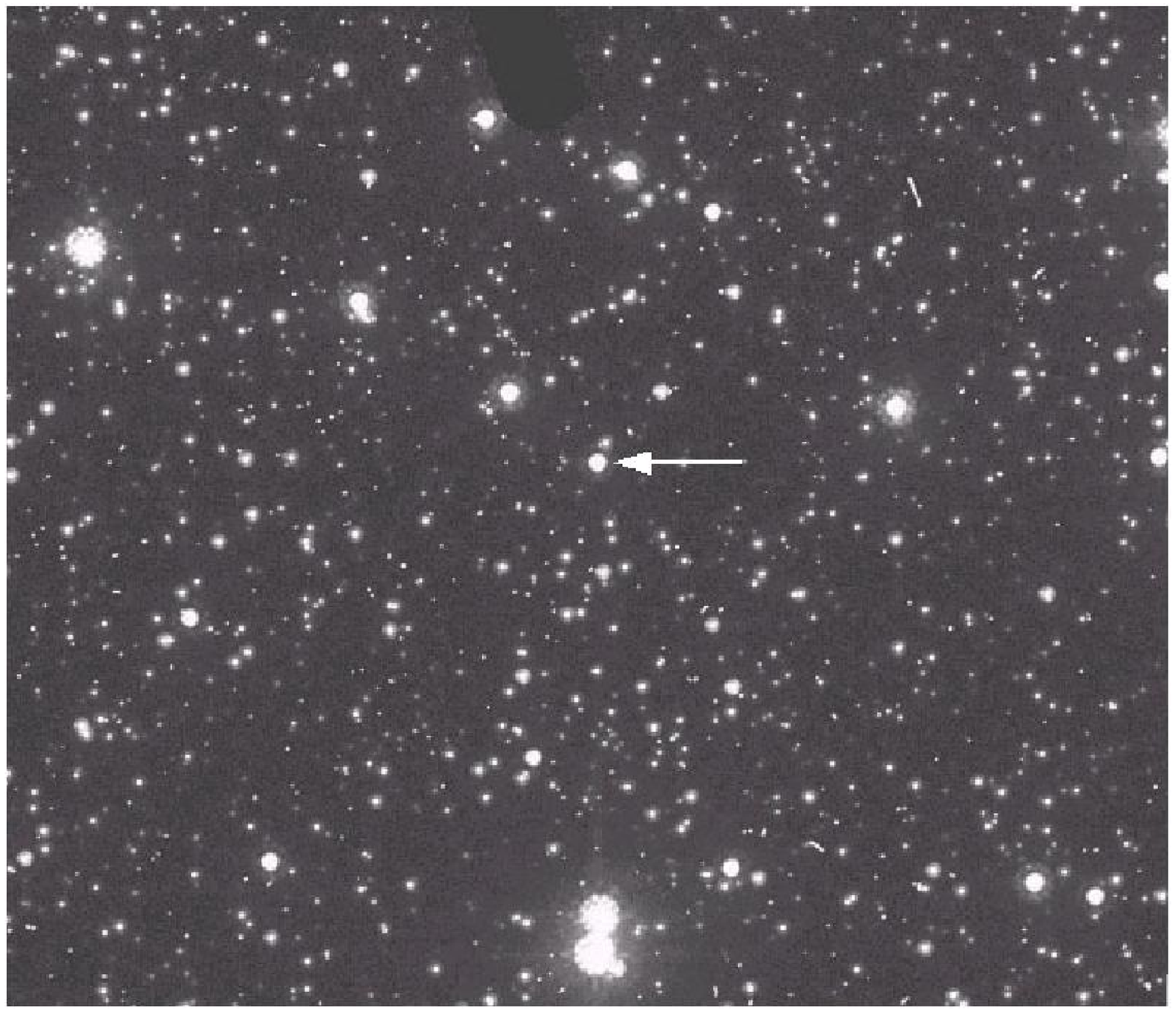}} &
\scalebox{1.8354}{\includegraphics{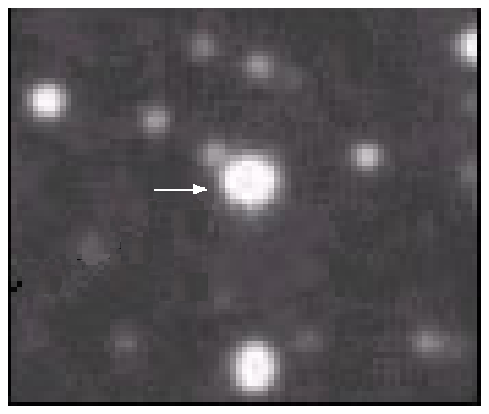}}
\end{tabular}
\caption{\label{fig:2x2}Left: HST image of MOA-33 (arrowed) in the F814W passband 39 days after peak magnification. This includes (slightly magnified) light from the  source star, light from the binary partner of the source star (see below), and also direct light from the members of the binary star comprising the lens (presumably fainter). Future high resolution images should be able to show the lens and source as resolved objects. The field of view is $26 \times 29$ arcsec. Right: MDM field in the I-band shown to the same scale but at high magnification.}
\end{center}
\end{figure}

HST images of MOA-33 were taken in the F435W, F555W and F814W filters with the ACS/HRC at 39 and 102 days following peak amplification.  Astrometry using stacked best-quality images allowed an identification of the MOA-33 source star in the HST images, see Figure~\ref{fig:2x2}. This identification was confirmed by aperture photometry of the HST images which showed a decline in luminosity of this star between the two observation epochs.

\setlength{\tabcolsep}{3pt}

\begin{figure}
\psfrag{xlabel1}{\LARGE{$HST_{\rm{F555W}}$}}
\psfrag{xlabel2}{\LARGE{$HST_{\rm{F814W}}$}}
\psfrag{ylabel1}{\LARGE{$V_{\rm{OGLE}}$}}
\psfrag{ylabel2}{\LARGE{$I_{\rm{OGLE}}$}}

\begin{center}
\scalebox{0.8}{\includegraphics{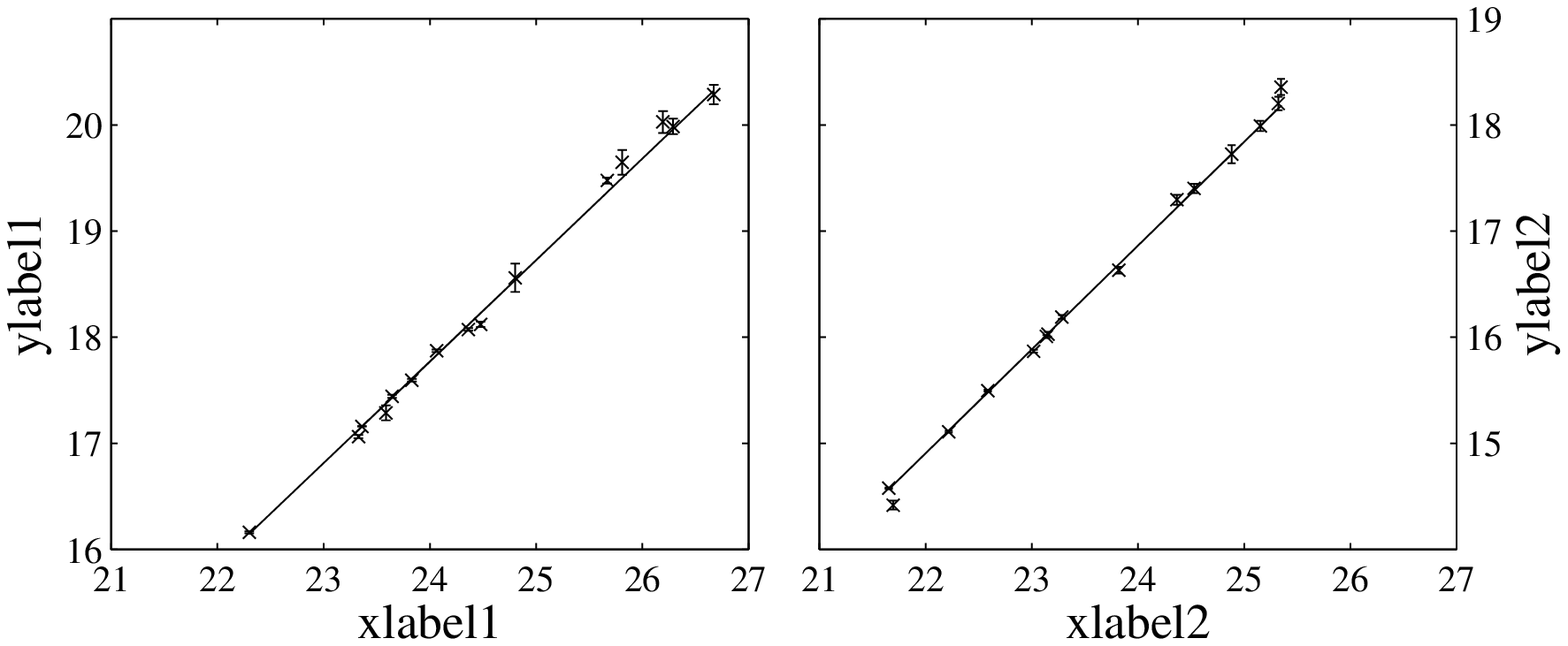}}
\caption[]{Conversion of HST instrumental magnitudes to OGLE apparent magnitudes.}
\label{fig:hst2ogle}
\end{center}
\end{figure}

Fifteen isolated stars on the HST images were identified in the OGLE catalogue of bulge stars (Udalski~\etal, 2002). Aperture photometry was performed to correlate HST instrumental magnitudes to OGLE $V$ and $I$ bands, see Figure~\ref{fig:hst2ogle}:

\begin{eqnarray}
V_{\rm{OGLE}} = (0.955 \pm 0.005)HST_{\rm{F555W}} + (-5.2 \pm 0.1) \label{eq:Vogle}\\
I_{\rm{OGLE}} = (0.978 \pm 0.005)HST_{\rm{F814W}} + (-6.6 \pm 0.1) \label{eq:Iogle}
\end{eqnarray}

\begin{figure}
\psfrag{xlabel}{$V_{OGLE} - I_{OGLE}$}
\psfrag{ylabel}{$I_{OGLE}$}
\begin{center}

\scalebox{0.7}{\includegraphics{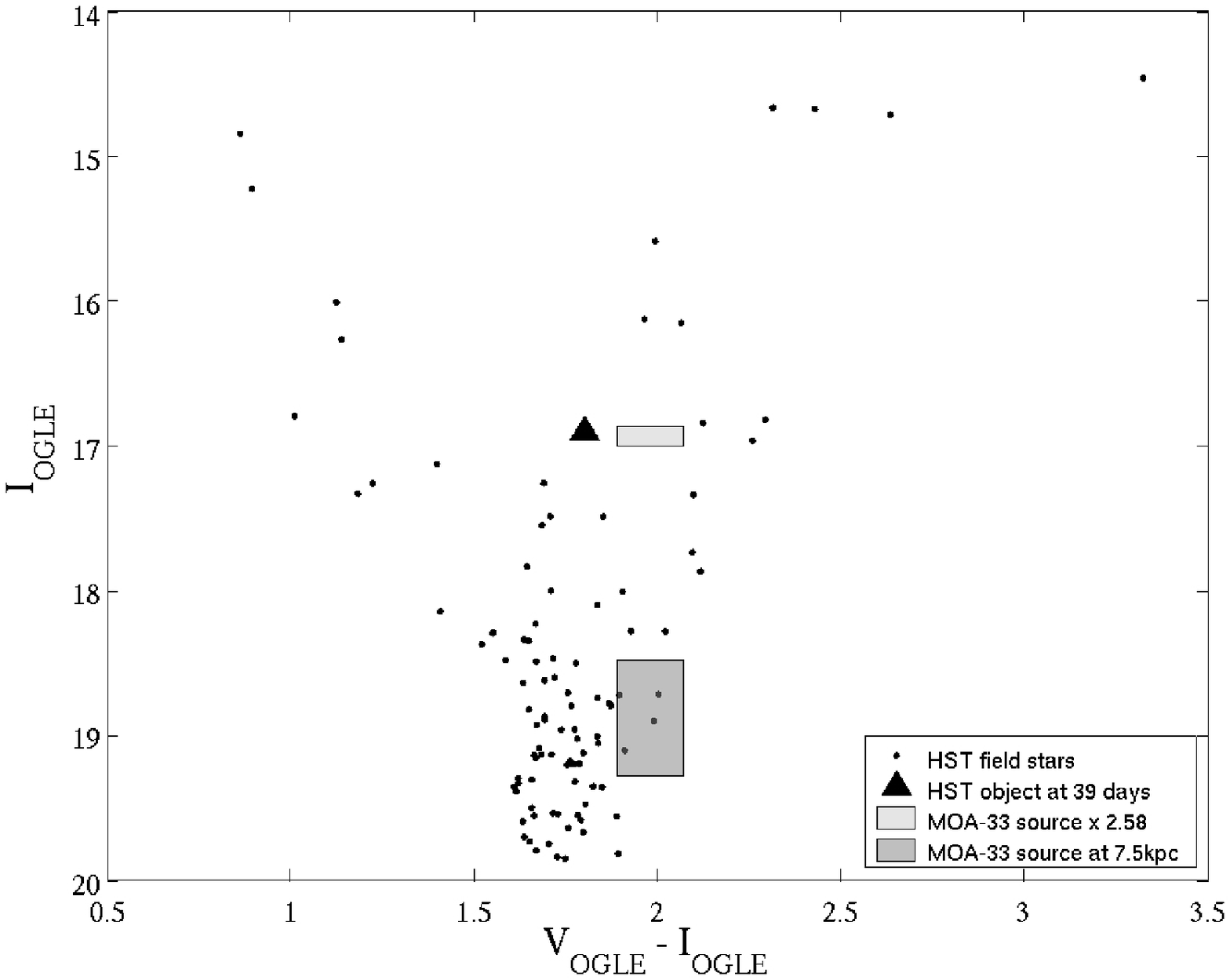}}

\caption[]{CMD of 100 isolated stars on HST images taken during the first observation epoch (at $t_{0} + 39$ days). The triangle denotes the HST object at the position of the microlensing event. This includes light from the magnified source star, light from an unmagnified binary partner of the source star, direct light from the binary lens, and light from any other possible stars along the line of sight. The dark grey rectangle denotes the position of the source star at unit magnification displaced to the common distance of the CMD stars (7.5 kpc). The light grey rectangle denotes the expected position of the HST object assuming the source was composed of a binary of two similar stars, one magnified by 1.58, and the other unmagnified (magnification = 1).}

\label{fig:moa33cmd}
\end{center}
\end{figure}

Figure~\ref{fig:moa33cmd} shows the CMD of 100 stars from the HST images converted to OGLE magnitudes via equations~\ref{eq:Vogle} and \ref{eq:Iogle}. The position of the source star of MOA-33 (unmagnified) is shown by the dark grey rectangle. The reddened values of  $I$ and $V - I$ were taken from Abe \etal\ (2003), but the former value was modified to allow for the fact that the comparison stars on the CMD are, on average, at the distance of the  bulge, whereas the source star of MOA-33 is at 4 - 6 kpc (Section 2). The dark grey rectangle corresponds to the source star displaced to 7.5 kpc. Its location on the CMD relative to the other stars is consistent with an F8 - G2  turn-off main-sequence star.

The expected magnitude change of the source star between the two HST observation epochs was $\Delta m = 0.44$ according to the fitted microlensing light curve for the event. However, the changes in $V$ and $B$ band magnitudes, $\Delta V =0.23 \pm 0.02$ and $\Delta B = 0.18 \pm 0.02$ were significantly less than this, suggesting the source for the event was a member of a binary comprised of two similar stars, only one of which was lensed.  Unfortunately, an $I$-band exposure of MOA-33 was made by the HST at only the first of the above epochs, i.e. 39 days after the peak magnification of the event, so no information is available on $\Delta I$.

Aperture photometry was carried out on the HST object at the 39 day epoch, when the magnification due to microlensing was 1.58. This is shown by the triangle in Figure~\ref{fig:moa33cmd}. Also shown in the figure (light grey rectangle) is its expected location assuming it is comprised of two similar stars, one magnified by 1.58, and the other unmagnified (i.e. magnification = 1). The approximate confluence of the two locations supports the above interpretation of the source star of MOA-33 as a member of a binary composed of two similar stars, only one of which was lensed. This assumes the contribution of direct light from the lens was small.

The binarity of the source star raises an interesting question. Our analysis has implicitly assumed that the motion of the source star during the microlensing event was rectilinear when this was clearly not the case. However, the critical microlensing period involves the caustic crossing only, and for this the duration was 15.6 hours. This is considerably less than the typical period of binary stars. We thus expect that this effect (the so-called ``xallarap" effect) may be safely neglected in MOA-33. It is, of course, for a similar reason that we may ignore the acceleration of the Earth (the ``parallax" effect) during the event.

\section{Source star shape modelling}
A total of 11 physical parameters and 8 flux normalisation parameters were required to model the light curve with a circular source star (Abe \etal, 2003)\footnote{The Wise data were not used in this analysis because they could not significantly affect the oblateness measurement.}. A source star which is non-spherical due to rotation would have an elliptical profile. To model an elliptical source star profile, two additional parameters were required: the source profile eccentricity, $e$, and attitude angle, $\beta$, with respect to the source star direction of motion. The eccentricity is defined as $e = \sqrt{1 - (b/a)^{2}}$, where $a$ and $b$ are the semi-major and semi-minor axes of the projected elliptical profile. These parameters are shown schematically in the inset of Figure~\ref{fig:caustic}. We note that it is not possible to determine the inclination of the poles of the source star with respect to the line of sight. This implies that we can only set a lower limit on the source oblateness.

Event MOA-33 was modelled using light curves generated by the inverse ray-shooting technique (Wambsganss, 1997) implemented on a cluster computer.  The best fitting light curves were found by minimising \chisq\ using a Monte Carlo Markov Chain technique (Rattenbury \etal, 2002). To set limits on the source shape the values of $e$ and $\beta$ were fixed on a 2D grid and all parameters were allowed to vary in order to minimise \chisq.  The source star limb-darkening was modelled using 

\begin{equation}
\frac{I_{\lambda}(\theta)}{I_{\lambda}(0)} = 1- c_{\lambda}(1-\cos \theta) - d_{\lambda}(1 - \sqrt{\cos \theta})
\end{equation}
where $c_{\lambda}$ and $d_{\lambda}$ are the filter-dependent limb-darkening coefficients and $\theta$ is the angle of the emitted light with respect to the line of sight. If $r_{\rm{s}}$ defines the (elliptical) source star profile and $r$ is the distance measured from the centre of the source star profile then the isophotes are concentric ellipses where $\cos \theta = \sqrt{1 - (r/r_{\rm{s}})^{2}}$. In general, a more complete model of the rotating source star would include the gravity-darkening effect of von Zeipel (1924), where the isophotes are not necessarily concentric ellipses. We note however that the von Zeipel theorem of gravity darkening is only valid for higher temperature stars (Claret 1998, Claret 2000) and would therefore be inapplicable here. While this correction was not made in this current work, we note that the inverse-ray shooting algorithm should be suitable for including this effect.

\section{Results}
\begin{figure}
\begin{center}
\psfrag{xlabel}{\LARGE{$\beta$}}
\psfrag{ylabel}{\LARGE{$e$}}
\scalebox{0.7}{\includegraphics{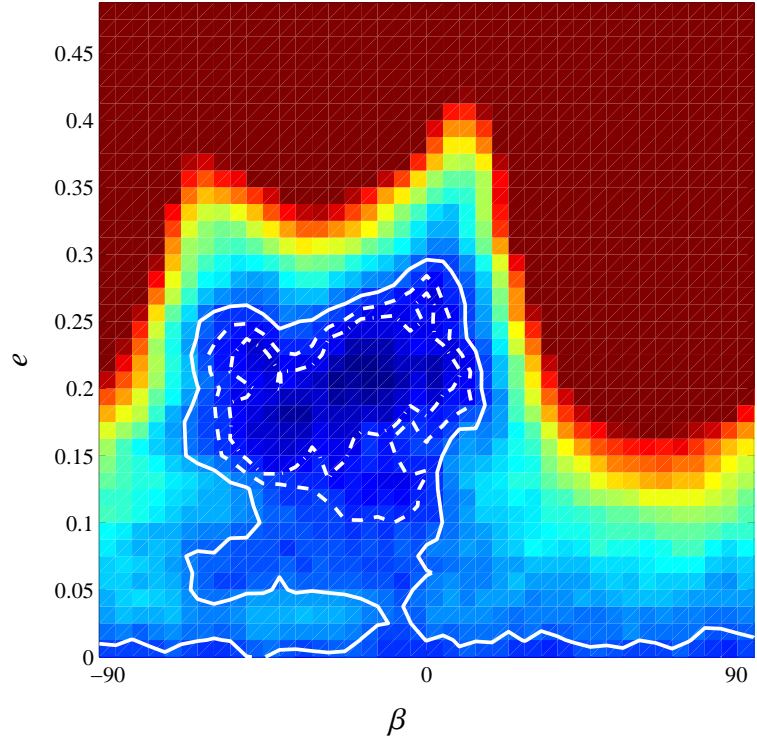}}
\caption[]{$\Delta \chi^{2}$ values from replacing the circular source profile used by Abe \etal\ (2003) with an elliptical source star profile. Limb darkening was held fixed at the values found by Abe \etal\ (2003), but all other fitting parameters are allowed to float, using a Markov Chain Monte Carlo model fitting algorithm. The colour code varies from dark blue~($\sim -20$) to light blue~($\sim30$), cyan~($\sim50$), green~($\sim60$), yellow~($\sim70$) and red~($\gtrsim90$). The contours are the $99\%$ (solid), $95\%$ (dashed) and $68\%$ (dot-dash) confidence limits.}
\label{fig:chisq}
\end{center}
\end{figure}

The \chisq\ minimisation  process was initially carried out with all 11 physical parameters introduced by  Abe \etal\ (2003) allowed to float over a coarse grid of values for the source shape parameters $e$ and $\beta$. The 8 flux parameters were also allowed to vary. It was found that the limb-darkening coefficients $c_{\lambda}$ and $d_{\lambda}$ so obtained did not vary significantly from the values obtained by Abe \etal\ (2003) for a circular source star. Subsequent \chisq\ minimisation was carried out for 40 values of $e$ ranging from $0$ to $0.485$ and 41 values of $\beta$ ranging from $-90^{\circ}$ to $90^{\circ}$ with the limb-darkening coefficients $c_{\lambda}$ and $d_{\lambda}$ held fixed at values found by Abe \etal\ (2003), allowing all other parameters to vary. The process was repeated six times to eliminate statistical artifacts. Figure~\ref{fig:chisq} shows the \chisq\ map so obtained.

The 99, 95 and 68\% confidence-level contours (corresponding to $\Delta \chi^{2} = 9.0, 4.0$ and $1.0$ respectively) in Figure~\ref{fig:chisq} show  regions in the $e-\beta$ parameter space that cannot be ruled out at these confidence levels. The 95\% contour yields $e = 0.17^{+0.11}_{-0.07}$ and $\beta = -24^{\circ} \pm 40^{\circ}$. The former value corresonds to $b/a = 0.98^{+0.01}_{-0.02}$.

Our treatment did not take account of rotation of the binary lens. During the caustic crossing shown in Figure~\ref{fig:caustic}, the lens will have rotated about an unknown axis through $1^{\circ}$ to $2^{\circ}$ (Abe~\etal, 2003), corresponding to an approximately 3\% undetermined change in the configuration of the binary lens components. We allow for this by a 3\% systematic uncertainty\footnote{In the case of face-on rotation, a linear relation between the errors is assured, because rotation through $n$ degrees merely rotates the caustic through $n$ degrees. In view of this, linearity should therefore be approximately valid in general.} added in quadrature in $b/a$. Thus $b/a = 0.98^{+0.02}_{-0.04}$, and therefore $e = 0.17\pm0.17$.

\section{Discussion}
The source star profile in a microlensing event may be non-circular for a number of reasons. The source star may possibly be rotating rapidly, for example, or it may be a member of a contact binary. We modelled the source star profile as an ellipse, with eccentricity, $e$, varying from 0 to 0.485. We have noted previously that it is not possible to infer the orientation of the source rotation axis. All estimates of source oblateness must therefore be considered lower limits. Figure~\ref{fig:chisq} shows some evidence of numerical noise, for instance, the expected degeneracy for all values of beta at $e = 0$ is not apparant everywhere: the 99\% contour is seen to touch the $e = 0$ line. Increasing the number of process iterations would remove this and other examples of numerical noise.

Table~\ref{tab:compare} compares the fitted lens and source parameter values for circular and elliptical source star profiles. We note that the lens parameter values for an elliptical source star model are not significantly different to those for a circular source profile. This suggests that any improvement in the model fit is due to the inclusion of an elliptical source star profile. However this analysis is not statistically preferred over the analysis of (Abe \etal, 2003). The elliptical source star profile that best fits the data has a profile that is only slightly oblate, and is moderately inclined to the projected motion.
 
\begin{table}
\begin{center}
\begin{tabular}{llll}
\hline \hline
Parameter & & \multicolumn{2}{c}{Source star profile}  \\
& & Circular & Elliptical\\
\hline
Impact parameter & $u_{\rm{min}}$ & $(4.29\pm0.09)\times 10^{-4}$ & $4.285 \times 10^{-4}$\\
Centre time & $t_{0}$ & HJD $2452460.496\pm 0.004$ & 2452460.496\\
Einstein time &  $t_{\rm{E}}$ & $(50.7 \pm 1.0)$ day & 50.73\\
Lens separation &  $d$ & $0.11\pm 0.01$ & 0.112\\ 
Lens mass ratio & $\frac{m_{2}}{m_{1}}$ & $0.54\pm0.20$ &0.54\\
Position angle & $\theta$ & $112^{\circ}\pm 2^{\circ}$& $112^{\circ}$\\
Eccentricity & $e$ & 0.0 & $e = 0.17\pm 0.17$\\ 
\hline
\end{tabular}
\caption{\label{tab:compare}Comparison of the lens and source parameters for a circular source star (Abe \etal, 2003)  and an oblate source star (this work). 
}
\end{center}

\end{table}

The colour of the source star is  $B-V = 0.57 \pm 0.04$ (Section 2). The maximum observed rotation speed for such a star  is approximately $20 \, \textrm{kms$^{-1}$}$ (Wolff \& Simon, 1997; Reiners \& Schmitt, 2003). The eccentricity of a source star profile given the Roche model is 
\begin{equation}
e^{2} = 1 - \left(\frac{1}{1 + \omega^{2} / 2} \right)^{2}
\end{equation}
where $\omega = \rm{v}_{\rm{rot}} / \rm{v}_{\rm{c}}$, $\rm{v}_{\rm{c}}$ is the critical velocity $\rm{v}_{\rm{c}} = \sqrt{{GM}/{R_{\rm{eq}}}}$, $\rm{v}_{\rm{rot}}$ is the speed of rotation (Pelupessy \etal, 2000) and $R_{\rm{eq}}$  is the equatorial radius of the star. We thus expect  $e\lesssim 0.05$ for MOA-33 , fully consistent with the measurement. 

It is possible that the star is crossing the Hertzsprung gap and is evolved from a highly rotating main-sequence early F star, which can rotate at speeds $\sim50 - 100 \, \textrm{kms$^{-1}$}$ (Kraft, 1967). However Endal \& Sofia (1979) show that a $1.5M_{\odot}$ star with an initial rotation velocity of $50 \, \textrm{kms}^{-1}$ would have a rotation speed of $\sim15 \, \textrm{kms}^{-1}$ at $B-V = 0.6$, assuming no internal angular momentum transport. Assuming rapid internal angular momentum transport, then a star with an initial rotation velocity of $65 \, \textrm{kms}^{-1}$ would have slowed to $\sim20  \, \textrm{kms}^{-1}$ at $B-V = 0.6$. The assumption that the maximum rotation velocity of the star is $20 \, \textrm{kms$^{-1}$}$ is therefore reasonable even if the star began its life rapidly rotating. In addition, the probability that the star is crossing the Hertzsprung gap is low compared to the probability that the star is on the main sequence.

\subsection{Expected frequency of similar events}
The results described here demonstrate how oblateness measurements can be made using caustic crossing microlensing events. It should be possible to make oblateness measurements of the source star in similar events, where the source star is straddled by two caustic lines at some time during an event. 

In order to determine (to first order) how likely such events are, we computed the fraction of binary lens events for which the source star is located between two caustic lines at least once during the event. We define $D_{\rm{cc}}$ as the distance the source profile travels between two caustic lines. The determination of source oblateness in the manner of this work is possible if  $D_{\rm{cc}}$ is of the order of the source star radius. As a guide, the distance between the caustic lines in event MOA-33 is $D_{\rm{cc, moa33}} \simeq 0.44R_{\rm{s}}$, where $R_{\rm{s}}$ is the MOA-33 source star radius.

$1.2\times10^{6}$ binary lens events were simulated with binary mass ratio and separation randomly drawn from log-uniform distributions. The minimum impact parameter and source star track angle values were drawn from uniform distributions. A solar radius source star at distance $D_{\rm{S}} = 8$ kpc and lens mass of $0.3M_{\odot}$ at $D_{\rm{L}}= 6$ kpc were assumed. Oblateness measurements will be optimised for $0 \lesssim f = D_{\rm{cc}}/R_{\rm{s}} \lesssim 1$.  

The fraction of binary lens systems that showed a caustic crossing signature was 0.094. The fraction of binary lens caustic crossing events for which $0 \leq f \leq 1$ was found to be $0.024$. The probability that a binary lens system will include a caustic crossing event that can constrain the source star shape is therefore $\simeq 2.3\times10^{-3}$. Assuming that a binary lens fraction of $0.5$, we would expect the fraction of microlensing events that have caustic crossing signatures which can constrain the source star shape to be $\sim 1.1\times 10^{-3}$. The OGLE and MOA microlensing collaborations have observed $\mathcal{O}(10^{3})$ microlensing events to date. The expected number of such events is therefore similar to the number actually observed, i.e. one. We note that these simulations are only good to first order, as we implicitly assume that all caustic crossing events are observed, and with sufficiently high accuracy and frequency to allow measurements of the source shape.

Measurements of the shapes of other distant stars would be valuable. We note that future microlensing  experiments can be expected to discover more events similar to MOA-33. The planned Microlensing Planet Finder space-based mission (Bennett \etal, 2004), for example, would offer an excellent opportunity for finding several such events, and measuring the shapes of several distant stars of various types.

\section{Conclusion}
These results constrain the projected eccentricity of the  MOA-33 source star profile to be $e = 0.17\pm 0.17$ and its corresponding oblateness to be $a/b = 1.02^{+0.04}_{-0.02}$. The source star type as determined by colour photometry of ground and HST images suggests a slow rotator star, which is consistent with this measurement.

The accuracy of the oblateness measurement reported here is equivalent to an angular resolution of $0.04R_{\odot} / D_{\rm{S}}$, where $D_{\rm{S}}$ is the distance to the source star. Assuming  $D_{\rm{S}} = (5 \pm 1)$ kpc (Section 2), the effective angular resolution of these measurements is $\simeq 0.04 \, \mu\rm{arcsec}$.

\begin{table}
\begin{center}
\begin{tabular}{lllll}
\hline \hline
Star & $m_{\rm{v}}$ & Spectral type & distance (pc) & Oblateness ($a/b$)\\
\hline
Altair & 0.8 & A7 IV-V & 5 & $1.140\pm0.029$\\
Achernar & 0.5 & B3 Vpe & 39 & $1.56\pm0.05$\\
MOA-33 & 19.9 & F8 - G2 V & $\sim5000$ & $1.02^{+0.04}_{-0.02}$\\ 
\hline

\end{tabular}
\caption{\label{tab:others}Comparison of the  MOA-33 source oblateness with the recent optical interferometry results for Achernar (Domiciano de Souza \etal, 2003), and Altair (van Belle \etal, 2001). }
\end{center}

\end{table}

In comparing these results with those reported by the optical interferometry teams for Achernar (Domiciano de Souza \etal, 2003) and Altair (van Belle \etal, 2001), see Table~\ref{tab:others},  it is important to note that these stars were already known to be fast rotators, and therefore expected to present elliptical profiles. While it is impossible to perform a similar preselection of targets in microlensing, theoretical simulations of binary lens caustic crossing events suggest that a fraction, $\sim1.1\times 10^{-3}$, of all microlensing events will be able to constrain the source oblateness, given sufficiently well-sampled observations. It is therefore possible that further events similar to MOA-33 will be observed and additional stellar profiles of distant stars obtained.

\section{Acknowledgements}
The authors thank the EROS collaboration for making their data available for this analysis. The MOA project is supported by the Marsden Fund of New Zealand, the Ministry of Education, Culture, Sports, Science and Technology (MEXT) of Japan, and the Japan Society for the Promotion of Science (JSPS). AG acknowledges support by NASA through Hubble Fellowship grant \#HST-HF-01158.01-A awarded by STScI, which is operated by AURA, Inc., for NASA, under contract NAS 5-26555. NJR thanks the Department of Computer Science, The University of Auckland for making computer resources available for this work.

\end{document}